\documentclass[%
reprint,
amsmath,amssymb,
pra,superscriptaddress
]{revtex4-1}
\usepackage[utf8]{inputenc}
\usepackage{graphicx}% Include figure files
\usepackage{dcolumn}% Align table columns on decimal point
\usepackage{bm}% bold math
\usepackage{mathrsfs}
\usepackage{amsfonts}
\usepackage{physics}
\usepackage{color}
\usepackage{xcolor}

\usepackage[english]{babel}
\usepackage{url}
\usepackage{bm}
\usepackage{hyperref}
\definecolor{darkblue}{rgb}{0,0,0.5}
\hypersetup{
colorlinks=true,
linkcolor=black,
filecolor=blue,
citecolor=darkblue,  
urlcolor=black,
}

\begin{document}

\title{Quantum-Enhanced Fiber-Optic Gyroscopes Using Quadrature Squeezing and Continuous Variable Entanglement}

\author{Michael R. Grace}
\affiliation{James C. Wyant College of Optical Sciences, University of Arizona, Tucson, AZ 85721, USA}
\author{Christos N. Gagatsos}
\affiliation{James C. Wyant College of Optical Sciences, University of Arizona, Tucson, AZ 85721, USA}
\author{Quntao Zhuang}
\affiliation{Department of Electrical and Computer Engineering, University of Arizona, Tucson, AZ 85721, USA}
\affiliation{James C. Wyant College of Optical Sciences, University of Arizona, Tucson, AZ 85721, USA}
\author{Saikat Guha}
\affiliation{James C. Wyant College of Optical Sciences, University of Arizona, Tucson, AZ 85721, USA}

\begin{abstract}
	We evaluate the fundamental performance of a fiber-optic gyroscope (FOG) design that is enhanced by the injection of quantum-optical squeezed vacuum. In the presence of fiber loss, we compute the maximum attainable enhancement over a classical, laser-driven FOG in terms of the rotation estimator variance from a homodyne measurement. We find that currently realizable amounts of single-mode squeezing are sufficient to access the maximum quantitative improvement, but that this gain in rotation sensitivity is limited to a marginal constant factor. We then propose an entanglement-enhanced FOG design that segments a fixed amount of available fiber into multiple fiber interferometers and feeds this sensor array with multi-mode-entangled squeezed vacuum. Our design raises the maximum improvement in sensitivity to an appreciable factor of $e\approx2.718$.
\end{abstract}

\maketitle

\section{Introduction}
\label{sec:Introduction}
Quantum-enhanced sensing~\cite{Pirandola2018,Giovannetti2011,Degen2017} is a rapidly growing field and one of the near-future tangible quantum technologies. In the domain of optical sensing, quantum-enhanced systems utilize nonclassical states of light and/or nonclassical optical detection schemes to improve the quantifiable performance of various sensing tasks. Among these nonclassical optical effects, squeezing and entanglement are two key backbones that are known to produce major performance enhancements in many emerging optical sensing and imaging applications~\cite{Abadie2011,Abbott2016,Lloyd2008,Tan2008,Giovannetti2001,Pirandola2011,Guha2013,Wasilevski2010}.

Quadrature-squeezed light is a well-established non-classical resource dating to pioneering works of Caves, Shapiro, Yuen and others ~\cite{Caves1981,Bondurant1984,Walls1983,Yuen1979}. Squeezing refers to a controllable phenomenon where quantum uncertainty is allocated asymmetrically between two non-commuting observables (e.g., the orthogonal quadratures of the complex field amplitude for an optical mode) while obeying the fundamental uncertainty relationship that lower bounds the product of their variances~\cite{Weedbrook2012,Adesso2014}. Squeezed vacuum (SV) can be generated using spontaneous parametric down conversion (SPDC), and experimental squeezing capability is steadily improving: vacuum with quantum noise reduction of $10$--$15$ dB in a given quadrature has been achieved in recent years~\cite{Vahlbruch2016,Schonbeck2018}. Quadrature squeezing is now routinely utilized in physical experiments relating to quantum information and sensing, most notably in the famous Laser Interferometer Gravitational-Wave Observatory (LIGO)~\cite{Abadie2011,Abbott2016}.

On the other hand, quantum entanglement, i.e., correlations that are stronger than and cannot be described by classical probability theory, is still being explored for enhancement of optical sensing systems. Recent theoretical and experimental works have established that entanglement can enable superior precision for quantitative sensing tasks including target detection~\cite{Lloyd2008,Tan2008}, positioning~\cite{Giovannetti2001}, digital memory readout~\cite{Pirandola2011,Guha2013}, and magnetometry~\cite{Wasilevski2010}.
%Surprisingly, this advantage can survive even when entanglement is broken by optical loss and/or noise, as exemplified by the quantum illumination scheme for target detection ~\cite{Tan2008,Guha2009,Barzanjeh2015,shapiro2019quantum}.

More recently, entanglement has been shown to give an advantage in measuring global features of a signal using a sensor network, a topic termed ``distributed quantum sensing"~\cite{Zhuang2018,Ge2018,Proctor2018}. One such idea is a continuous-variable (CV) distributed quantum sensing protocol~\cite{Zhuang2018,Zhuang2020} that generates a multi-mode entangled optical probe by splitting single-mode squeezed vacuum using a passive beamsplitter network. CV distributed quantum sensing, which has been experimentally validated~\cite{Guo2019,Xia2019}, is especially attractive from the experimental standpoint due to robustness against optical losses. The core idea in Ref.~\cite{Zhuang2018} has led to applications such as optical beam-displacement tracking~\cite{Qi2018} and quantum-enhanced machine learning for optical-sensor-based signal classification~\cite{Zhuang2019a} and has potential for other sensing contexts. 

The fiber-optic gyroscope (FOG) is a high precision, compact solution for precision navigation in GPS-denied environments, ultraprecise platform stabilization, and other inertial sensing applications~\cite{Lefevre2014f}. FOGs leverage the Sagnac principle~\cite{Arditty1981}, i.e., that light traveling along one branch of a rotating interferometer will undergo a detectable phase shift with respect to the light in the other branch. Like all fiber-based sensors, optical loss is a significant factor in FOG performance. Recent developments include hollow core fiber coils~\cite{Digonnet2016}, integrated optical sources and homodyne detectors, and laser-driven FOGs that bypass the conventional requirement for temporally incoherent sources~\cite{Lloyd2013,Lloyd2013a}. In addition, quantum-enhancement has recently been considered to boost rotation sensitivity ~\cite{Haus1991,Mehmet2010,Stevenson2015,Kok2017,Luo2017,Fink2019,Yu2019,Xu2019}, and preliminary experimental works on quantum-enhanced FOGs have demonstrated improvements via injection of squeezed vacuum~\cite{Mehmet2010} and entangled NOON states~\cite{Fink2019}.

In this paper, we first establish the fundamental sensitivity achievable by the squeezing-enhanced FOG design proposed in Ref.~\cite{Mehmet2010}, for which single-mode SV is injected to boost sensitivity. Our quantifying figure of merit for sensitivity is the estimator variance from sensing a small rotation. Then, inspired by recent theoretical work on distributed quantum sensing~\cite{Zhuang2018,Ge2018,Proctor2018}, we propose and analyze the fundamental performance of a novel CV entanglement-enhanced FOG design that uses a stacked array of multiple identical interferometers, where one part of an entangled optical state is distributed to each. 

The main results of our investigation are the following:
\begin{itemize}
	\item {\textbf{Squeezing-Enhanced FOG}}---In Sec.~\ref{sec:ClassicalFOG} we analyze a {\em classical FOG} (termed \textit{Design C}), i.e., a fiber Sagnac interferometer read by a laser, as a baseline sensor design (Fig.~\ref{fig:ClassicalQuantumFOG}). In Sec.~\ref{sec:SqueezingEnhancedFOG}, we calculate the rotation sensitivity achievable with the \textit{squeezing-enhanced FOG} (\textit{Design S}) reported in Ref.~\cite{Mehmet2010}. Ignoring losses, Design S achieves a Heisenberg-limited scaling advantage in sensitivity ~\cite{Pirandola2018,Giovannetti2011,Degen2017}; with fiber loss, the quantum advantage falls to a constant factor that depends on the amount of squeezing. Optimizing Design S over fiber length, we find that squeezing beyond 10--15 dB yields diminishing returns in sensitivity (Fig.~\ref{fig:HeisenbergSQL}), so currently existing technology could access the quantum advantage. However, we show in Sec.~\ref{sec:ClassicalComparison} that Design S provides at best a small improvement factor of $1.196$ over Design C, capping its practical potential.
	
	\item {\textbf{CV Entanglement-Enhanced FOG}}---In Sec. \ref{sec:EntanglementEnhancedFOG}, we propose a \textit{CV entanglement-enhanced FOG} (\textit{Design E}), in which single-mode SV is split into an $M$-mode entangled state and injected into $M$ parallel fiber interferometers (Fig. \ref{fig:distributedFOG}). We compare its sensitivity to that of a \textit{product-state squeezing-enhanced FOG} (\textit{Design P}), which is equivalent to $M$ parallel copies of Design S each with an independent squeezer, and a baseline \textit{distributed classical FOG} (\textit{Design D}), which is $M$ independent copies of Design C. Assuming a constant dB/km of fiber loss, we evaluate these designs in two contexts:
	
	1. \underline{Unconstrained Total Fiber Length}: With optimized total fiber lengths, Design E surpasses Design S in sensitivity by a factor of $M$, mirroring the $M$-fold improvement of Design D over Design C under a peak power constraint per fiber. Design P improves upon Design D in sensitivity but is outperformed by Design E; Design P only achieves the same linear scaling with $M$ as Design E if both its number of squeezers and its total squeezing power also scale with $M$ (Fig. \ref{fig:BarChart}). Our resource-efficient Design E, with just a single squeezer, leverages both distributed sensing and quantum sensing. Still, Sec. \ref{sec:ClassicalComparison} shows that the improvement of Design E over Design D is again limited to $1.196$ (Fig. \ref{fig:ClassicalComparison}).
	
	2. \underline{Constrained Total Fiber Length}: If the total fiber length is held fixed under a weight or payload requirement, Design E again outperforms Designs D and P when $M$ is optimized for each (Fig. \ref{fig:Fixed_Fiber_Length}). 
	%In particular, the greatest performance gap between Designs E and P occurs in the regime of 5--15 dB of squeezing when both designs are given the same squeezed light energy.
	Unlike with unconstrained fiber length, Design E attains a notable sensitivity improvement over Design D that is independent of both the fiber loss (dB/km) and the total fiber length. We find in Sec. \ref{sec:ClassicalComparison} a maximum improvement factor of $e \approx 2.718$ in the limit of infinite squeezing (Fig. \ref{fig:ClassicalComparison}). Most of this benefit is attainable with 10--15 dB of squeezing, so our Design E could provide an appreciable real- world sensitivity enhancement in the near future.
	
\end{itemize}
\section{Classical FOG}
\label{sec:ClassicalFOG}

\subsection{Sagnac Interferometry} 
\begin{figure}[htbp]
	\centering
	\includegraphics[width=\linewidth]{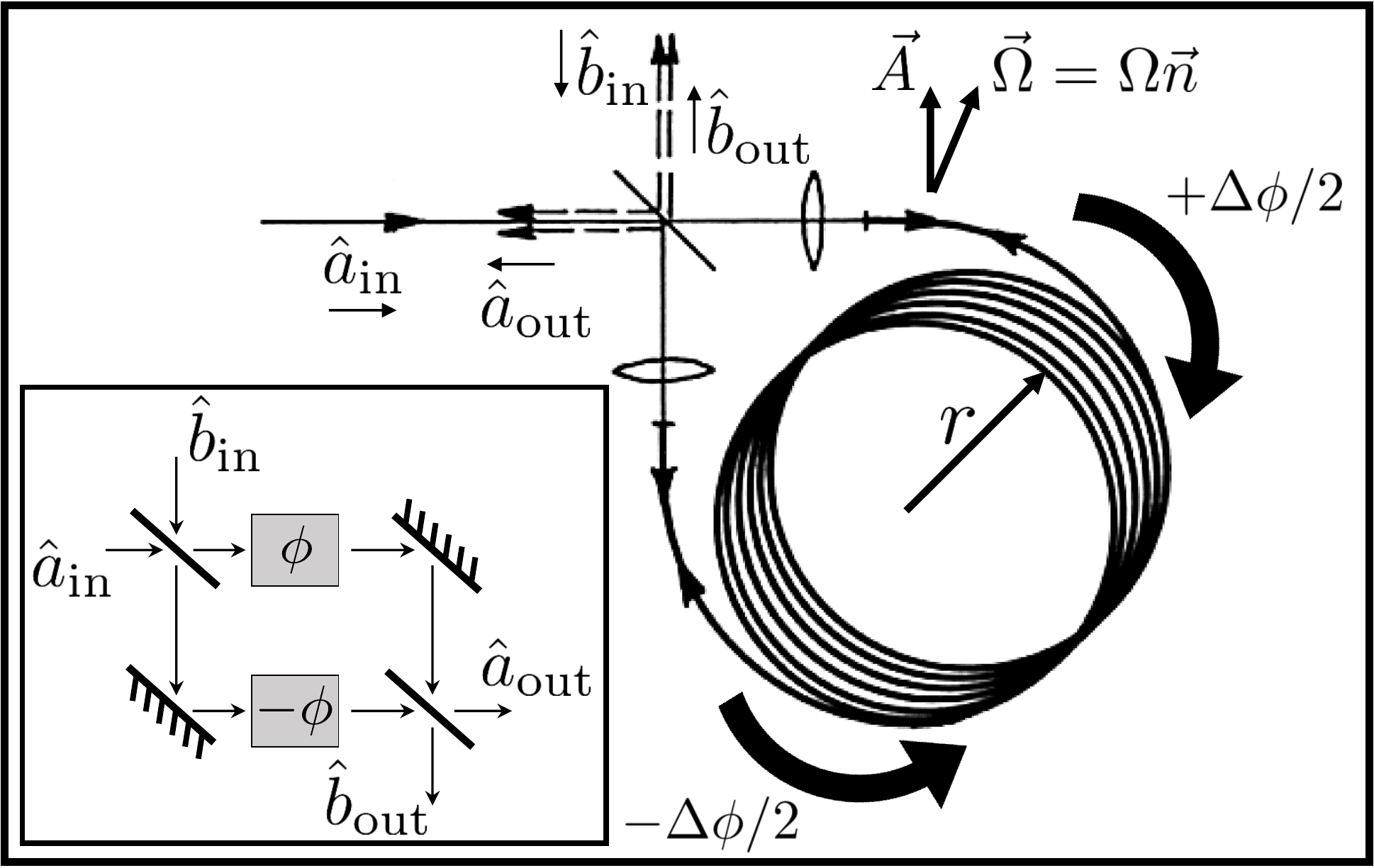}
	\caption{Fiber-optic gyroscope (FOG). Input and output mode pairs, identified by field operators $\hat{a}_{\textrm{in}}$, $\hat{b}_{\textrm{in}}$, $\hat{a}_{\textrm{out}}$ and $\hat{b}_{\textrm{out}}$, are spatially co-located at the external ports of the coupling beamsplitter. Adapted from~\cite{Lefevre2014f}. Inset: Mach-Zehnder interferometer with conjugate phases $\pm\phi$. The input and output modes can be conceptually mapped to those of the FOG.}
	\label{fig:FOG}
\end{figure}

A Sagnac interferometer measures angular velocity via the optical path delay induced in counter-propagating paths around a rotating loop (Fig.~\ref{fig:FOG}). For a laser-driven FOG, the resulting relative phase shift is given by~\cite{Arditty1982}
\begin{equation}
\Delta\phi=\frac{4\omega m\vec{A}\cdot \vec{\Omega}}{c^2},
\label{eq:sagnac}
\end{equation}
where $\omega$ is the optical center frequency of the laser, $m$ is the number of fiber loops in the coil, $\vec{A}$ is the directed area of a single fiber loop, $\vec{\Omega}=\Omega\vec{n}$ is the directed angular velocity of rotation with magnitude $\Omega\in\mathbb{R}$ and unit-norm direction vector $\vec{n}$, and $c$ is the speed of light in vacuum. If $L$ is the fiber length of a circular coil and $r$ is its radius, then the integer number of fiber loops in the coil $m=L/(2\pi r)$ linearly scales the total effective area of the gyroscope. ``The indices of refraction of the core and the cladding, the phase or group velocities, the dispersion of the medium or of the waveguide have no influence" on the non-reciprocal phase delay accumulated by the counter-propagating beams~\cite{Arditty1982}. Eq.~\ref{eq:sagnac} is valid to first order in $r\Omega/c$, corresponding to the standard regime of slow rotations compared with the speed of light~\cite{Arditty1981}.

In optical fiber-based technologies, a key source of optical loss is the transmission loss accumulated over the length of the fiber due to scattering from imperfections and/or evanescent coupling. We use $\eta$ to denote the effective transmissivity of each branch around the Sagnac interferometer. For a FOG with a fiber length of $L$ km, this transmissivity is given by $\eta=10^{-b L/10}$, where $b$ is a wavelength-dependent fiber loss coefficient with a typical value of $b\approx0.5$ dB loss per km at 1550 nm. 

\subsection{Conjugate Phase Sensing with Quantum Optics}
A Sagnac interferometer can be conceptually equated to a phase conjugate Mach-Zehnder interferometer (Fig. \ref{fig:FOG} inset) where $\pm\phi\equiv \pm\Delta\phi/2$ are the conjugate phases accumulated on the two optical paths due to the physical rotation of the sensor. In a FOG setup, the two input ports and the two output ports of the conceptual Mach-Zehnder are spatially overlapped on a single beamsplitter (Fig. \ref{fig:FOG}) and can be separated using a Faraday rotator, for example~\cite{Mehmet2010}. In quantum terminology, we define ($\hat{a}_{\textrm{in}}$, $\hat{a}_{\textrm{out}}$) and ($\hat{b}_{\textrm{in}}$, $\hat{b}_{\textrm{out}}$) as pairs of quantum field (annihilation) operators of spatially overlapping optical modes with opposite propagation directions (Fig. \ref{fig:FOG}).

A homodyne measurement can be used to estimate the conjugate phase $\phi$ and subsequently the angular velocity $\Omega=2\phi/T$, where $T=4\omega L(\vec{A}\cdot\vec{n})/(2\pi r c^2)$. In Appendix \ref{apx:conjugage_phase_sensing} we derive the measurement statistics obtained by a quantum-noise-limited homodyne measurement performed along the imaginary quadrature of mode $\hat{b}_{\textrm{out}}$. Letting $\tilde{b}_{\textrm{out}}$ designate the output random variable from such a homodyne measurement, the mean of $\tilde{b}_{\textrm{out}}$ is
\begin{equation}
	\begin{aligned}
	\expval*{\tilde{b}_{\textrm{out}}}=&\expval*{\textrm{Im}[\hat{b}_{\textrm{out}}]} \\
	=&
	\sqrt{\eta}\big(\sin(\phi)\expval*{\textrm{Re}[\hat{a}_{\textrm{in}}]}-\cos(\phi)\expval*{\textrm{Im}[\hat{b}_{\textrm{in}}]}\big),
	\end{aligned}
	\label{eq:homodyne_mean}
\end{equation}
and its variance is
\begin{equation}
	\begin{aligned}
		\expval*{\Delta\tilde{b}_{\textrm{out}}^2}=&\expval*{\Delta\textrm{Im}[\hat{b}_{\textrm{out}}]^2}\\
		=&
		\eta\big(\sin(\phi)^2\expval*{\Delta\textrm{Re}[\hat{a}_{\textrm{in}}]^2} \\
		&+\cos(\phi)^2\expval*{\Delta\textrm{Im}[\hat{b}_{\textrm{in}}]^2}\big)+\frac{1-\eta}{4}.
		\end{aligned}
	\label{eq:homodyne_variance}
\end{equation}
For a given FOG modality(Design $X$), we will identify an unbiased estimator $\tilde{\Omega}_X$ with a mean equal to the true value of the parameter of interest $\Omega$. The estimator variance will depend on $\expval*{\tilde{b}_{\textrm{out}}}$ and $\expval*{\Delta\tilde{b}_{\textrm{out}}^2}$, where the expectation values of the corresponding quantum field operators are taken over the quantum states $\ket{\psi_a}$ and $\ket{\psi_b}$ of the optical modes at the two input ports. 

\subsection{Classical FOG Sensitivity}
\label{sec:classicalFOG}
\begin{figure}[htbp]
	\centering
	\includegraphics[width=\linewidth]{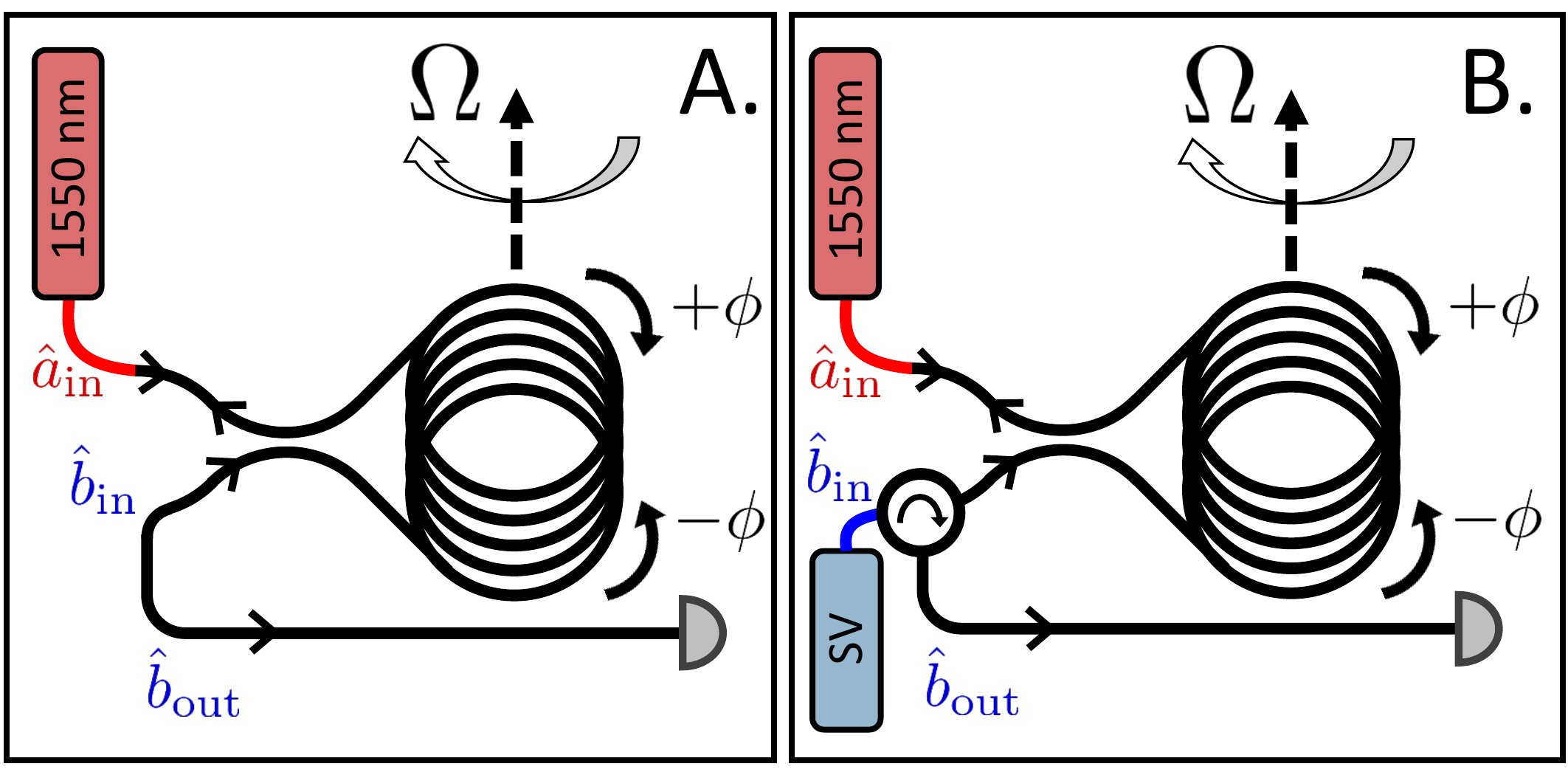}
	\caption{A. Classical FOG (Design C) with coherent state in the $\hat{a}_{\textrm{in}}$ mode and vacuum in the $\hat{b}_{\textrm{in}}$ mode followed by homodyne detection of the $\hat{b}_{\textrm{out}}$ mode. B. Squeezing-enhanced FOG (Design S) with single-mode squeezed vacuum (SV) injected at mode $\hat{b}_{\textrm{in}}$. The $\hat{b}_{\textrm{out}}$ mode is coupled to the homodyne detector using a Faraday rotator~\cite{Mehmet2010}. The local oscillators used for homodyning are not shown.}
	\label{fig:ClassicalQuantumFOG}
\end{figure}

In a conventional laser-driven FOG setup (Design C), the conceptual input mode $\hat{a}_{\textrm{in}}$ is fed with a laser, while mode $\hat{b}_{\textrm{in}}$ is physically overlapped by the output mode $\hat{b}_{\textrm{out}}$ and has no input (Fig. \ref{fig:ClassicalQuantumFOG}A). The joint input across the two ports can be modeled using the two-mode quantum state $\ket{\psi_a}\ket{\psi_b}=\ket{\alpha}\ket{0}$, a tensor product between a coherent state with mean photon number $N_v\equiv\expval*{\hat{a}_{\textrm{in}}^{\dagger}\hat{a}_{\textrm{in}}}{\psi_a}=\alpha^2$ and a vacuum state ($\alpha\in\mathbb{R}$ without loss of generality). In the small rotation regime ($\phi\ll 1$), the homodyne measurement statistics become
\begin{equation}
	\begin{aligned}
		\expval*{\tilde{b}_{\textrm{out}}}&=\sqrt{\eta}\big(\sin(\phi)\alpha-\cos(\phi)\,0\big)\approx \phi\sqrt{\eta}\alpha \\
		\expval*{\Delta\tilde{b}_{\textrm{out}}^2}&=\eta\big(\sin(\phi)^2\frac{1}{4}+\cos(\phi)^2\frac{1}{4}\big)+\frac{1-\eta}{4}=\frac{1}{4}.
	\end{aligned}
	\label{eq:ClassicalHomodyne}
\end{equation}
Notice that the mean is scaled by $\sqrt{\eta}$ while the variance arises from quantum noise and does not depend on loss. Thus, $\tilde{\Omega}_{\textrm{C}}=2\tilde{b}_{\textrm{out}}/(T\sqrt{\eta}\alpha)$ is an unbiased estimator of the unknown parameter $\Omega$. The sensitivity to rotation is quantified by the estimator variance:
\begin{equation}
	\expval*{\Delta\tilde{\Omega}_{\textrm{C}}^2}\approx\frac{1}{T^2\eta \alpha^2}=\frac{1}{T^2\eta N_v}.
	\label{eq:ClassicalVariance}
\end{equation} 
This estimator variance scales inversely with the total probe energy $N=N_v$, an example of the so-called standard quantum limit (SQL) that sets the best possible scaling with input energy for the variance of any measurement using purely classical probes and detectors. In terms of the fiber properties, the estimator variance is
\begin{equation}
\expval*{\Delta\tilde{\Omega}_{\textrm{C}}^2}\approx\frac{1}{V^{-2}N_v}\frac{1}{L^210^{-b L/10}},
\label{eq:DesignCvar}
\end{equation}
where we define $V=L/T=\pi r c^2/\big(2\omega (\vec{A}\cdot\vec{n})\big)$ to clarify the dependence of sensitivity on fiber length.

\section{Squeezing-Enhanced FOG}
\label{sec:SqueezingEnhancedFOG}
In this section, we compare the rotation sensitivity attained by classical and squeezing-enhanced FOGs as a function of input energy. The squeezing-enhanced FOG utilizes single-mode SV, for which squeezing is typically quantified in units of dB of quantum noise reduction in the squeezed observable with respect to vacuum. Denoting the ``dB of squeezing" by $\sigma$, the mean photon number of an ideal single-mode, quadrature-squeezed vacuum state is $N_s=\sinh^2(\sigma\ln(10)/20)$. 

\subsection{Squeezing-Enhanced FOG Sensitivity}
\label{sec:quantumFOG}
Our setup for a squeezing-enhanced FOG (Design S) is shown in Fig.~\ref{fig:ClassicalQuantumFOG}B, which is functionally equivalent to the design demonstrated in Ref.~\cite{Mehmet2010}. The vacuum input of Design C is replaced by vacuum squeezed along its imaginary quadrature, i.e. $\ket{\psi_b}=\ket{0;\mu,-\nu}$, where the parameters $\mu,\nu\in\mathbb{R}$ obey $\mu^2-\nu^2=1$ and the mean photon number of the SV is $N_s\equiv\expval*{\hat{b}_{\textrm{in}}^{\dagger}\hat{b}_{\textrm{in}}}{\psi_b}=\nu^2$~\cite{Yuen1979}. Considering the same ideal homodyne measurement as in Design C in the presence of loss and in the $\phi\ll 1$ regime,
\begin{equation}
\begin{aligned}
\expval*{\tilde{b}_{\textrm{out}}}&=\sqrt{\eta}\big(\sin(\phi)\alpha-\cos(\phi)\,0\big)\approx\phi\sqrt{\eta}\alpha \\
\expval*{\Delta\tilde{b}_{\textrm{out}}^2}&=\eta\bigg(\sin(\phi)^2\frac{1}{4}+\cos(\phi)^2\frac{(\mu-\nu)^2}{4}\bigg)+\frac{1-\eta}{4}\\
&\approx\frac{\eta(\mu-\nu)^2+1-\eta}{4}.
\label{eq:homodyne2}
\end{aligned}
\end{equation}
The unbiased estimator $\tilde{\Omega}_{\textrm{S}}=2\tilde{b}_{\textrm{out}}/(T\sqrt{\eta}\alpha)$  has variance
\begin{equation}
\begin{aligned}
\expval*{\Delta\tilde{\Omega}_{\textrm{S}}^2}&\approx\frac{\eta(\mu-\nu)^2+1-\eta}{T^2\eta\alpha^2} \\
&=\frac{1}{T^2\eta N_v}\bigg(\frac{\eta}{(\sqrt{1+N_s}+\sqrt{N_s})^2}+1-\eta\bigg).
\end{aligned}
\label{eq:SqueezingEnhancedVariance}
\end{equation} 

\begin{figure}[tbp]
	\centering
	\includegraphics[width=\linewidth]{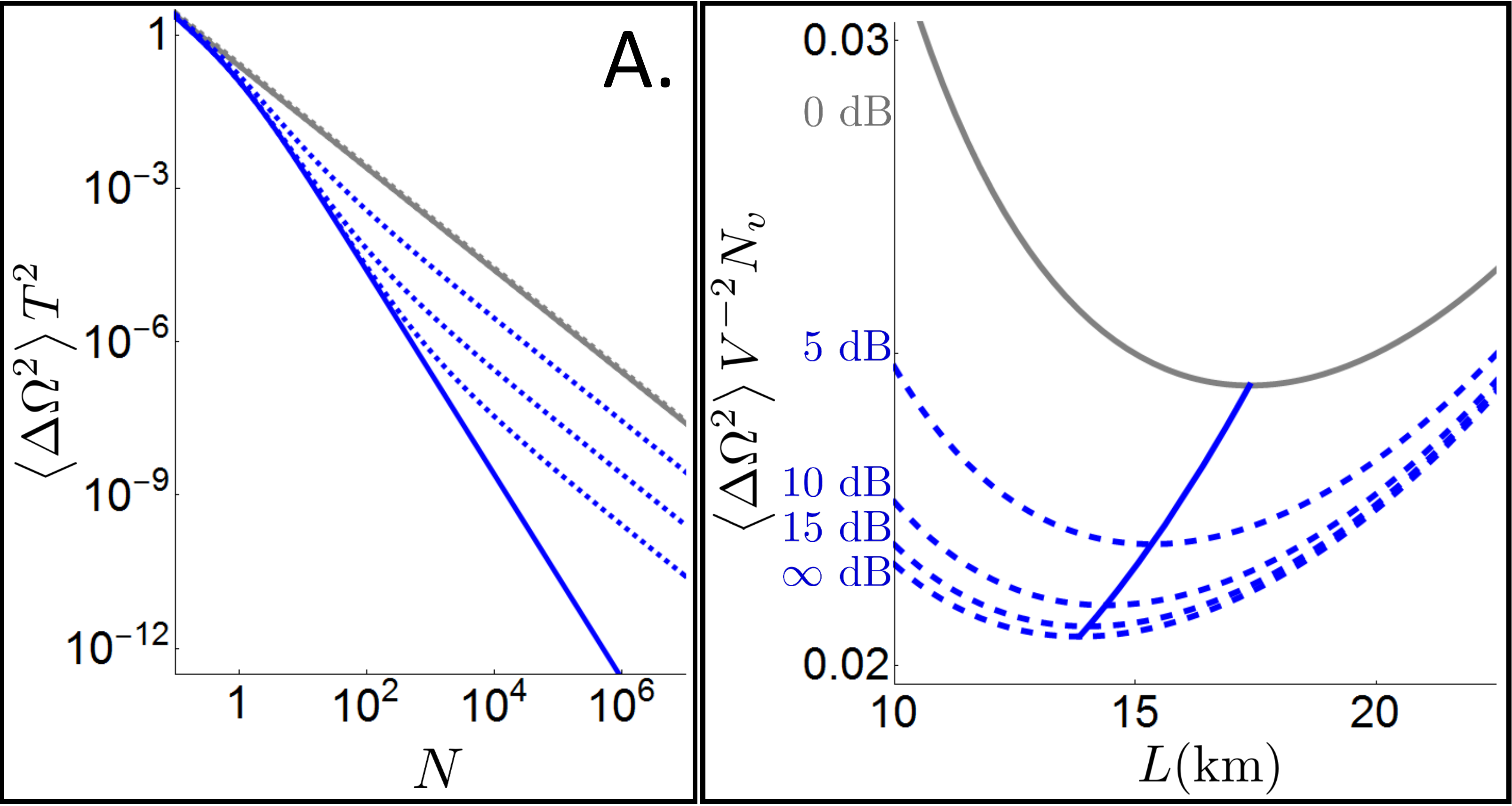}
	\caption{A. Normalized estimator variance for Design C (gray) and Design S (blue) for different fiber transmissivities $\eta$. Without loss, Design S exhibits Heisenberg limited scaling, whereas loss results in SQL scaling. Solid lines: $\eta=1$. Dotted lines from top to bottom: $\eta=0.9$, $\eta=0.99$, and $\eta=0.999$. B. Fiber length dependence of FOG rotation estimator variance with independent laser power and squeezing constraints and with $b=0.5$ km/dB of fiber loss. Gray line: Design C. Blue dashed lines from top to bottom: Design S with 5 dB squeezing, 10 dB, 15 dB, $\infty$ dB. The descending solid blue parametric curve indicates the Design S estimator variance with optimized fiber lengths as the squeezing is increased.}
	\label{fig:HeisenbergSQL}
\end{figure}

If the total input mean photon number $N=N_v+N_s$ is constrained, the probe energy can be allocated between the laser and the SV to minimize the estimator variance. After some algebra, this minimum variance is found to be
\begin{equation}
\expval*{\Delta\tilde{\Omega}_{\textrm{S}}^2}\vert_{N_s=N_{s,\rm opt}}\approx\frac{2(1-\eta)^2}{T^2\eta(1+2(1-\eta)N-z)},
\label{eq:SqueezingEnhancedVariance2}
\end{equation} 
where $z=\sqrt{1+4\eta(1-\eta)N}$, and the corresponding SV mean photon number is $N_{s,\rm opt}=2\eta^2N^2/\big(1+z+2\eta N(2-\eta+z)\big)$. In the absence of loss, i.e. $\eta=1$, the minimized estimator variance becomes $\lim_{\eta\to1}\{\expval*{\Delta\tilde{\Omega}_{\textrm{S}}^2}\vert_{N_s=N_{s,\rm opt}}\}=1/\big(T^2N(N+1)\big)$. This inverse quadratic scaling with input energy, known as the Heisenberg limit (e.g., $\expval*{\Delta\tilde{\Omega}_{\textrm{S}}^2}\propto 1/N^2$), is a unique advantage of quantum sensing and is the best scaling allowed by physics for any sensing task~\cite{Pirandola2018,Giovannetti2011,Degen2017}. Fig.~\ref{fig:HeisenbergSQL}A shows the difference in scaling with input energy between the classical and squeezing-enhanced FOG if loss is ignored. However, in the realistic case where $\eta<1$, Heisenberg scaling is lost; for high input energy, the minimized estimator variance limits to $\lim_{N\to\infty}\{\expval*{\Delta\tilde{\Omega}_{\textrm{S}}^2}\vert_{N_s=N_{s,\rm opt}}\}=(1-\eta)/(T^2\eta N)$. In this case, SQL scaling with probe energy is retained for Design S (i.e., $\expval*{\Delta\tilde{\Omega}_{\textrm{S}}^2}\propto \expval*{\Delta\tilde{\Omega}_{\textrm{C}}^2}\propto1/N$), yielding a constant factor improvement over Design C (Fig.~\ref{fig:HeisenbergSQL}A). 

A more appropriate approach to the optimization of Design S is to independently constrain $N_v$, the mean photon number from the laser, and $N_s$, the mean photon number of the injected SV. This distinction accounts for the fact that in practice squeezed light is more challenging or energy inefficient to produce than classical laser light, such that energy in a squeezed vacuum state is regarded as more ``expensive" than energy in a coherent state. A full analysis constraining the total ``wallplug" power required to pump the SPDC and the laser would be insightful but is not included here. 

For a lossless ($\eta=1$) squeezing-enhanced FOG in the high squeezing regime ($N_s\gg 1$, or $\sigma\gg7.66$ dB), the rotation estimator variance (Eq. \ref{eq:SqueezingEnhancedVariance}) scales inversely with the product of the mean photon numbers from the two optical inputs (i.e., $\expval*{\Delta\tilde{\Omega}_{\textrm{S}}^2}\propto 1/N_vN_s$), an alternative formulation of the Heisenberg limit. As before, optical loss restores SQL scaling (i.e., $\expval*{\Delta\tilde{\Omega}_{\textrm{S}}^2}\propto \expval*{\Delta\tilde{\Omega}_{\textrm{C}}^2}\propto1/N_v$). In this case, squeezing-enhancement offers a constant factor improvement to sensitivity for a given $N_s$ that we quantify using the sensitivity ratio $R_{\textrm{S}}=\expval*{\Delta\tilde{\Omega}_{\textrm{S}}^2}/\expval*{\Delta\tilde{\Omega}_{\textrm{C}}^2}$, where a smaller ratio means a greater improvement in quantum-enhanced performance over the classical baseline. The best achievable constant factor improvement for a given transmissivity $\eta$ occurs with infinite squeezing and is easily calculated to be $\lim_{N_s\to\infty}\{R_{\textrm{S}}\}=1-\eta$. 

\subsection{Squeezing-Enhanced FOG Sensitivity with Optimized Fiber Length}
In Fig. \ref{fig:HeisenbergSQL}B, we plot the estimator variances for Designs C and S to compare their performance as a function of fiber length. Applying the fiber loss model, the estimator variance for Design S (Eq.~\ref{eq:SqueezingEnhancedVariance}) becomes
\begin{equation}
\begin{aligned}
\expval*{\Delta\tilde{\Omega}_{\textrm{S}}^2}\approx& \frac{1}{V^{-2}N_v}\frac{1}{L^2 10^{-bL/10}} \\
&\bigg(\frac{10^{-bL/10}}{(\sqrt{1+N_s}+\sqrt{N_s})^2}+1-10^{-bL/10}\bigg).
\end{aligned}
\label{eq:DesignSvar}
\end{equation} 
%For example, a typical, high-resolution FOG with 10 km of 0.5 dB/km fiber \cite{Lefevre2014f} has an overall transmissivity of $\eta\approx0.316$, such that the best constant factor reduction in estimate variance is $\lim_{N_s\to\infty}R_{\textrm{S}}=1-\eta\approx0.684$. 
The relative performance gap can apparently be widened simply by cutting the total fiber length to suppress loss. However, this lowers the effective area of the interferometer (Eq.~\ref{eq:sagnac}), reducing absolute sensitivity to rotation. 

Due to this balance, it is possible to find an optimal fiber length that minimizes the estimator variances for Designs C and S when there is no prior constraint on $L$. FOGs are often designed with shorter total fiber length than the optimal value in order to boost the dynamic range in $\Omega$ and reduce the size and weight of the device~\cite{Lefevre2014f}. Still, the optimal fiber length will yield the best possible sensitivity to small rotations, a desirable feature of a quantum-enhanced FOG. 

To minimize the estimator variance for Design C (Eq.~\ref{eq:DesignCvar}), we solve $\partial\expval*{\Delta\tilde{\Omega}_{\textrm{C}}^2}/\partial L=0$ and find the solution
\begin{equation}
L_{\textrm{opt}}=\frac{20}{\ln(10)b}\approx\frac{8.686}{b},
\label{eq:ClassicalLopt}
\end{equation}
which is the fiber length known to optimize the signal-to-noise ratio of a classical FOG given a dB/km loss specification of $b$~\cite{Lefevre2014f}. Assuming $b=0.5$ dB/km at 1550 nm, $L_{\textrm{opt}}\approx17.372$ km. The optimized estimator variance is
\begin{equation}
\expval*{\Delta \tilde{\Omega}_{\textrm{C}}^2}\vert_{L=L_\textrm{opt}}=
\frac{1}{V^{-2}N_v}\frac{e^2\ln(10)^2b^2}{400}.
\label{eq:ClassicalFOGvariance}
\end{equation}

For the minimized estimator variance of Design S (Eq.~\ref{eq:DesignSvar}), solving $\partial\expval*{\Delta \tilde{\Omega}_{\textrm{S}}^2}/\partial L=0$ yields the solution
\begin{equation}
L_{\textrm{opt}}=\frac{10(2+\Lambda(N_s))}{\ln(10)b}
\label{eq:Lopt2}
\end{equation} 
where
\begin{equation}
\Lambda(x)=W\bigg(\frac{4\big(x-\sqrt{x(1+x)}\big)}{e^2}\bigg)
\label{eq:Lambda}
\end{equation}
and $W(y)$ is the principal value of the Lambert W function. This corresponds to a minimized variance of
\begin{equation}
\expval*{\Delta \tilde{\Omega}_{\textrm{S}}^2}\vert_{L=L_\textrm{opt}}=
\frac{1}{V^{-2}N_v}\frac{e^{2+\Lambda(N_s)}\ln(10)^2b^2}{200(2+\Lambda(N_s))}.
\label{eq:SqueezingEnhancedFOGvariance}
\end{equation}

The parametric curve in Fig. \ref{fig:HeisenbergSQL}B shows that SV injection boosts the optimal FOG sensitivity while reducing the length of fiber needed for optimal performance. Furthermore, experimentally feasible single-mode squeezing (10--15 dB) achieves nearly the full constant-factor quantum advantage; more squeezing yields diminshing returns. This indicates that a fully realized squeezing-enhanced FOG is a near-term attainable technology.

\section{Entanglement-Enhanced FOG}
\label{sec:EntanglementEnhancedFOG}
In this section we consider distributed sensing, where multiple identical Sagnac interferometers  are simultaneously employed in parallel orientations on a single spool to measure the same physical rotation. We use $M$ to denote the number of fiber-optic interferometers being used. For distributed FOGs, it may be appropriate to constrain not the total mean photon number $N_v$ of the laser light, but the \textit{per-mode} mean photon number $n_v=N_v/M$, or equivalently a per-mode laser power constraint, to account for the peak power permitted by an optical fiber before the onset of adverse nonlinear effects~\cite{Saleh2007}. This constraint introduces an immediate benefit in sensitivity via an increased total laser power budget. We investigate the additional enhancement in sensitivity that is possible when squeezed light is injected into a distributed FOG. While CV entanglement does not provide a significant quantum advantage when the total fiber length is unconstrained, we find that it can yield an appreciable improvement under a total fiber length constraint.

\begin{figure*}[htb]
	\centering
	\includegraphics[width=\linewidth]{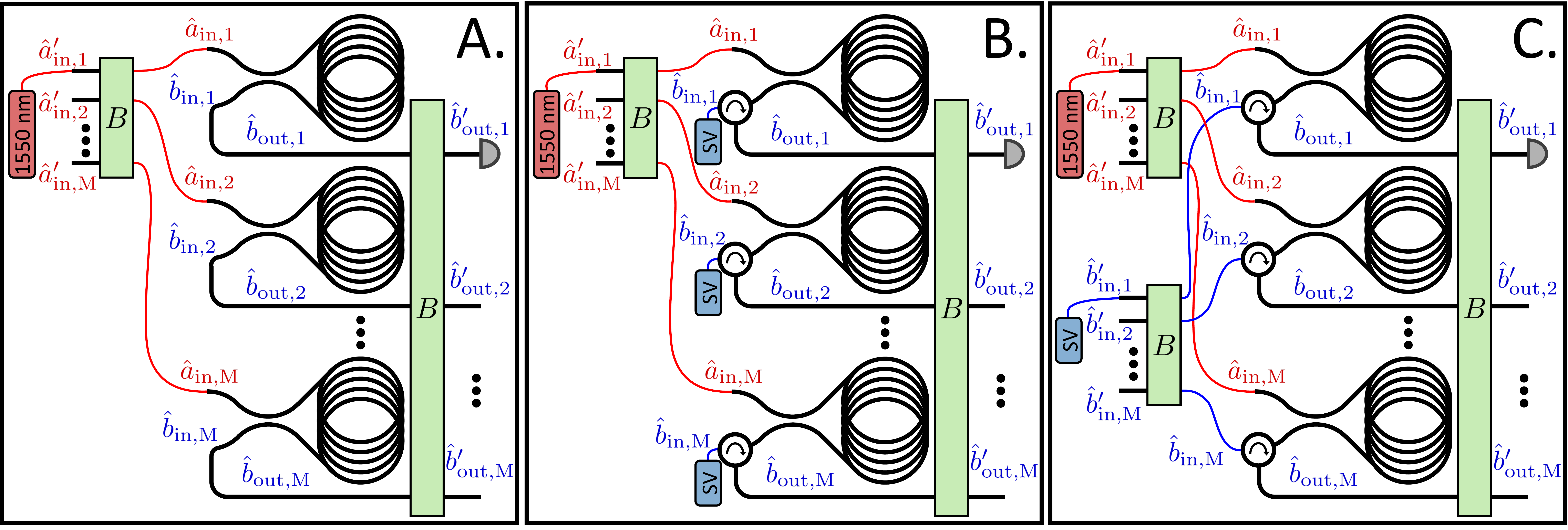}
	\caption{A. Distributed classical FOG (Design D) with each interferometer fed by the same laser. B. Product-state squeezing-enhanced FOG (Design P) with separable SV inputs generated using $M$ single-mode squeezers. C. Entanglement-enhanced FOG (Design E) with SV from one squeezer distributed to the $M$ interferometers. While the Sagnac loops are drawn as separated in space, in practice they could be mounted on the same fiber spool with independent fiber couplers.}
	\label{fig:distributedFOG}
\end{figure*}

\subsection{Distributed FOG Modalities}
\label{sec:distributedFOGmodalities}
%For a distributed sensing scheme, it may be appropriate to constrain not the total mean photon number $N_v$ and $N_s$ designated to laser light and squeezed radiation but the \textit{per mode} mean photon numbers $n_v$ and $n_s$. The laser power per mode constraint $n_v$ reflects a peak power permissible by an optical fiber before the onset of significant nonlinear effects. The squeezed vacuum power per mode $n_s$ corresponds to the practical limitation of single mode squeezing, which is a function of nonlinear crystal properties [?]. 

Fig. \ref{fig:distributedFOG} shows three distributed FOG configurations. In all three cases, the $M$ interferometers are fed by modes $\hat{a}_{\textrm{in},j}, j\in[1,M]$ in coherent states $\ket{\psi_{a,j}}=\ket*{\alpha/\sqrt{M}}$. These coherent states could be sourced from a single laser driving the coherent state $\ket{\psi_{a'}}=\ket{\alpha}$ in mode $\hat{a}'_{\textrm{in},1}$ with mean photon number $N_v\equiv\expval*{\hat{a}^{\prime\dagger}_{\textrm{in},1}\hat{a}'_{\textrm{in},1}}{\psi_{a'}}=\alpha^2$, where this laser light is equally split $M$ ways using a balanced beamsplitter array ($B$ in Fig.~\ref{fig:distributedFOG}A-C). An identical beamsplitter array recombines the $M$ symmetric output modes $\hat{b}_{\textrm{out},j}$, and a single homodyne measurement is performed on mode $\hat{b}'_{\textrm{out},1}$~\cite{Zhuang2019a}. The statistics of a quantum-noise-limited, lossless, imaginary-quadrature homodyne measurement output variable $\tilde{b}'_{\textrm{out},1}$ are calculated in Appendix~\ref{apx:distributed_conjugate_phase_sensing} for each of the three configurations.

A distributed classical FOG (Design D) is depicted in Fig.~\ref{fig:distributedFOG}A. The $\hat{b}_{\textrm{in},j}$ mode on each interferometer is in a vacuum state $\ket{\psi_b}=\ket{0}$, and the joint quantum state is $\ket{\psi_{a'}}\ket{\psi_b}^{\otimes M}$, with $\otimes M$ denoting a tensor product. Since coherent states and vacuum states remain pure through passive linear optical transformations such as the beamsplitter arrays $B$, the light effectively undergoes the same evolution as that of Design C, with homodyne output statistics given by Eq.~\ref{eq:ClassicalHomodyne}. If $\eta$ is the optical transmissivity of each identical interferometer, the variance of the unbiased estimator $\tilde{\Omega}_{\textrm{D}}=2\tilde{b}'_{\textrm{out},1}/(T\sqrt{\eta}\alpha)$ is given by
\begin{equation}
	\expval*{\Delta\tilde{\Omega}_{\textrm{D}}^2}\approx\frac{1}{T^2\eta \alpha^2}=\frac{1}{T^2\eta M n_v},
	\label{eq:DistributedVariance}
\end{equation} 
an $M$-fold improvement over a single interferometer (Eq.~\ref{eq:ClassicalVariance}) when the laser power per fiber $n_v$ is constrained.
 
In the simplest squeezing-enhanced, distributed FOG (Design P), identical SV states $\ket{\psi_{b}}=\ket{0;\mu,-\nu}$ are introduced independently into each of the $M$ interferometers (Fig.~\ref{fig:distributedFOG}B). The optical input to this system is again in the product state $\ket{\psi_{a'}}\ket{\psi_b}^{\otimes M}$, where the mean photon number of the SV injected into each interferometer is $n_s=N_s/M$ and the total mean photon number from squeezed light is $N_s\equiv\sum_{j=1}^M\expval*{\hat{b}_{\textrm{in},j}^{\dagger}\hat{b}_{\textrm{in},j}}{\psi_{b}}=M\nu^2$. The mean of the homodyne output variable $\tilde{b}^{\prime 2}_{\textrm{out},1}$ is the same as that of Design S (Eq.~\ref{eq:homodyne2}), and the variance is
\begin{equation}
	\begin{aligned}
		\expval*{\Delta\tilde{b}^{\prime 2}_{\textrm{out},1}}&=\eta\bigg(\frac{\sin(\phi)^2}{4}+\frac{\cos(\phi)^2}{M}\sum_{j=1}^M\frac{(\mu-\nu)^2}{4}\bigg)+\frac{1-\eta}{4}\\
		&\approx\frac{\eta(\mu-\nu)^2+1-\eta}{4}.
		\label{eq:ProductHomodyne}
	\end{aligned}
\end{equation}
Using the unbiased estimator $\tilde{\Omega}_{\textrm{P}}=2\tilde{b}'_{\textrm{out},1}/(T\sqrt{\eta}\alpha)$, 
\begin{equation}
\expval*{\Delta\tilde{\Omega}_{\textrm{P}}^2}\approx\frac{1}{T^2\eta M n_v}\bigg(\frac{\eta M}{(\sqrt{M+N_s}+\sqrt{N_s})^2}+1-\eta\bigg).
\label{eq:ProductVariance}
\end{equation} 

Finally, we propose a CV entanglement-enhanced FOG (Design E) that injects multi-mode CV entangled light into the $M$ interferometers (Fig. \ref{fig:distributedFOG}C). The multi-partite entangled state can be generated by mixing single-mode SV of mean photon number $N_s\equiv\expval*{\hat{b}_{\textrm{in},1}^{\prime\dagger}\hat{b}_{\textrm{in},1}^{\prime}}{\psi_{b'}}=\nu^2$ with $M-1$ vacuum modes on another  balanced beamsplitter array $B$~\cite{Zhuang2018}, the output of which is distributed to the $M$ interferometers. The resulting $\hat{b}_{\textrm{in},j}$ modes exhibit mutual quantum entanglement and cannot be written as a tensor product of quantum states. In Appendix \ref{apx:distributed_conjugate_phase_sensing} we show that $\hat{a}'_{\textrm{in},j}$ and $\hat{b}'_{\textrm{in},j}$ evolve into $\hat{a}'_{\textrm{out},j}$ and $\hat{b}'_{\textrm{out},j}$ via the exact same mathematical relationships as for Design S, so the mean and variance of the homodyne measurement output $\tilde{b}'_{\textrm{out},1}$ are given by expressions equivalent to Eqs. \ref{eq:homodyne_mean} and \ref{eq:homodyne_variance}. The unbiased estimator $\expval*{\Delta\tilde{\Omega}_{\textrm{E}}^2}=2\tilde{b}'_{\textrm{out},1}/(T\sqrt{\eta}\alpha)$ has a variance of
\begin{equation}
\expval*{\Delta\tilde{\Omega}_{\textrm{E}}^2}\approx\frac{1}{T^2\eta M n_v}\bigg(\frac{\eta }{(\sqrt{1+N_s}+\sqrt{N_s})^2}+1-\eta\bigg).
\label{eq:variance5}
\end{equation} 

\subsection{Quantum-Enhanced Distributed FOG Sensitivity with Unconstrained Total Fiber Length}
For Design D with $\eta=10^{-b L/10}$, minimizing the estimator variance (Eq.~\ref{eq:DistributedVariance}) over the total fiber length by solving $\partial\expval*{\Delta\tilde{\Omega}_{\textrm{D}}^2}/\partial L=0$ gives the optimal total length
\begin{equation}
L_{\textrm{opt}}=\frac{20M}{\ln(10)b}=\frac{8.686M}{b},
\label{eq:Lopt6}
\end{equation}
which equals the optimal total length for Design C (Eq.~\ref{eq:ClassicalLopt}) scaled by $M$. The minimized variance is then
\begin{equation}
\expval*{\Delta \tilde{\Omega}_{\textrm{D}}^2}\vert_{L=L_\textrm{opt}}=
\frac{1}{V^{-2}Mn_v}\frac{e^2\ln(10)^2b^2}{400},
\label{eq:Omega_DLopt}
\end{equation}
a factor of $M$ reduced from $\expval*{\Delta \tilde{\Omega}_{\textrm{C}}^2}\vert_{L=L_\textrm{opt}}$ (Eq.~\ref{eq:ClassicalFOGvariance}), since $n_v$ is equal to $N_v$ with a single interferometer.

Optimizing the total fiber length for Design E reveals a similar relationship with respect to Design S. Solving $\partial\expval*{\Delta\tilde{\Omega}_{\textrm{E}}^2}/\partial L=0$ using Eq.~\ref{eq:variance5} gives
\begin{equation}
L_{\textrm{opt}}=\frac{10(2+\Lambda(N_s))M}{\ln(10)b},
\label{eq:Lopt8}
\end{equation} 
the optimal fiber length for Design S (Eq.~\ref{eq:Lopt2}) scaled by M. This is evidence that Design E is the true distributed sensing analog for Design S. Likewise,
\begin{equation}
\expval*{\Delta \tilde{\Omega}_{\textrm{E}}^2}\vert_{L=L_\textrm{opt}}=
\frac{1}{V^{-2}Mn_v}\frac{e^{2+\Lambda(N_s)}\ln(10)^2b^2}{200(2+\Lambda(N_s))},
\label{eq:Omega_ELopt}
\end{equation}
equivalent to $(1/M)\expval*{\Delta\tilde{\Omega}_{\textrm{S}}^2}\vert_{L=L_\textrm{opt}}$ (Eq.~\ref{eq:SqueezingEnhancedFOGvariance}).

For Design P, the equation $\partial\expval*{\Delta\tilde{\Omega}_{\textrm{P}}^2}/\partial L=0$ can be solved using Eq.~\ref{eq:ProductVariance} to find
\begin{equation}
L_{\textrm{opt}}=\frac{10(2+\Lambda_M(N_s))M}{\ln(10)b},
\label{eq:Lopt7}
\end{equation}
where
\begin{equation}
\Lambda_M(x)=W\bigg(\frac{4\big(x-\sqrt{x(M+x)}\big)}{Me^2}\bigg).
\label{eq:Lambda_M}
\end{equation}
Note that $\Lambda_1(x)=\Lambda(x)$, since Designs S, P and E are identical for $M=1$. The optimized estimator variance is
\begin{equation}
\expval*{\Delta \tilde{\Omega}_{\textrm{P}}^2}\vert_{L=L_\textrm{opt}}=
\frac{1}{V^{-2}Mn_v}\frac{e^{2+\Lambda_M(N_s)}\ln(10)^2b^2}{200(2+\Lambda_M(N_s))},
\label{eq:FOGvariance7}
\end{equation} 
which has no simple relationship with $\expval*{\Delta\tilde{\Omega}_{\textrm{S}}^2}\vert_{L=L_\textrm{opt}}$.

\begin{figure}[htbp]
	\centering
	\includegraphics[width=\linewidth]{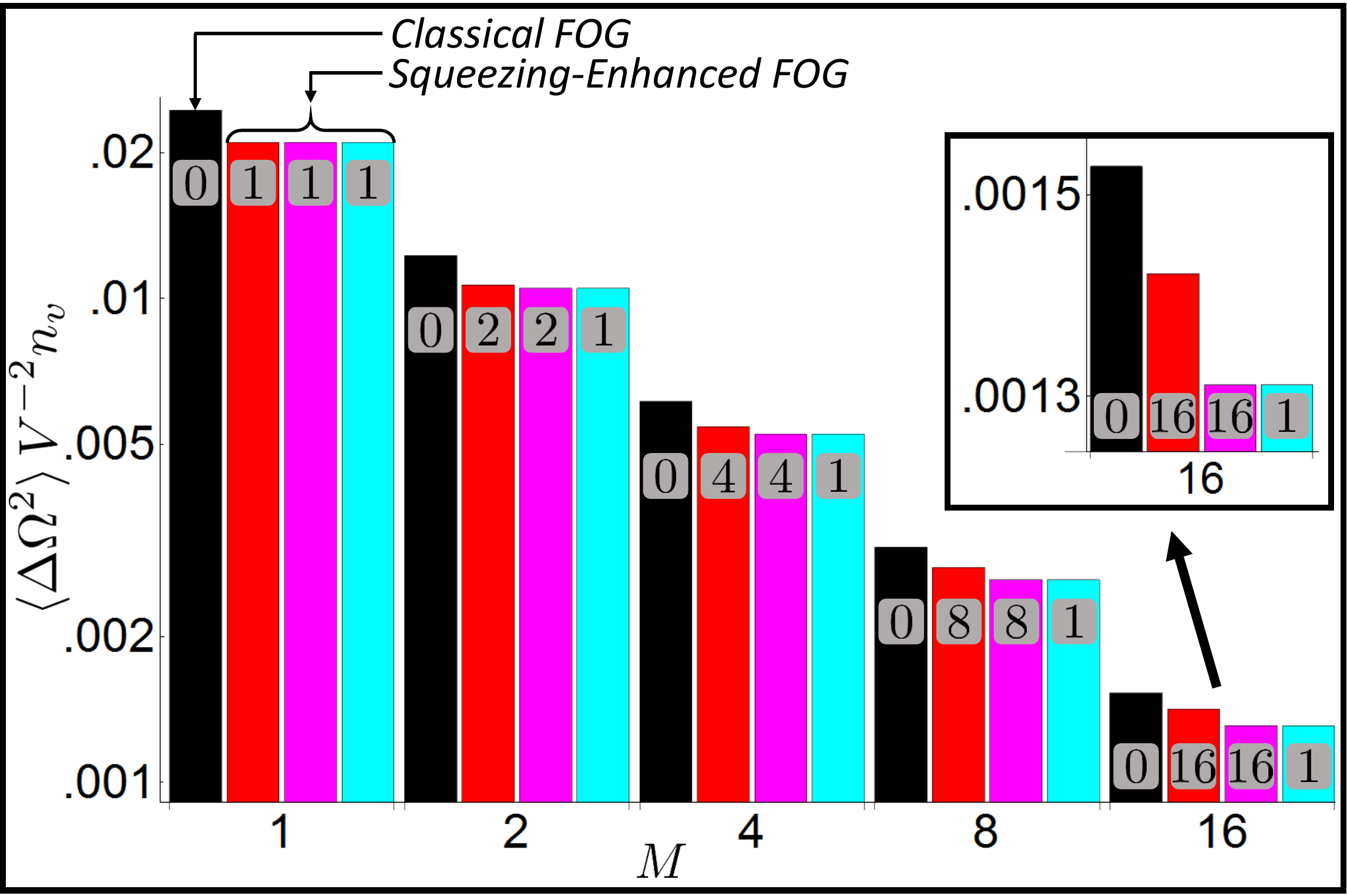}
	\caption{Normalized rotation estimator variance of distributed FOG modalities with optimized total fiber length and $b=0.5$ dB/km of loss. Black bars: Design D. Red bars: Design P with total squeezed-light energy equivalent to the energy from a single 10 dB squeezer. Magenta bars: Design P with $M$ 10 dB squeezers. Cyan bars: Design E using a single 10 dB squeezer. Numbers within bars indicate the number of single-mode squeezers required to implement the FOG design. Inset: magnified view of results for $M=16$.}
	\label{fig:BarChart}
\end{figure}

Figure \ref{fig:BarChart} compares the sensitivity of these three distributed FOG modalities when the total available fiber length $L$ is optimized for each case. Scaling the input laser power with $M$ yields an immediate advantage independent of squeezing, as seen for Design D (black bars). Given a fixed total energy from squeezed light $N_s$, Design E (cyan bars) increasingly outperforms Design P (red bars) as $M$ increases. Alternatively, if the \textit{per-mode} energy $n_s$ of each injected SV state is constrained due to a practical limitation on single-mode squeezing technology, Design P (magenta bars) matches the sensitivity of Design E (for which $N_s=n_s$) for all values of $M$. However, to match this performance Design P requires $M$ physical single-mode squeezers, each with the same pump energy as that of the one squeezer used for Design E (Fig.~ \ref{fig:distributedFOG}). This result showcases the value of entanglement-enhanced sensing in terms of resource efficiency: using CV quantum entanglement, a single SV source enables the same quantitative performance as that achieved by several of the same SV sources placed in parallel. Still, the additional improvements arising from injected SV for Designs P and E over Design D are only marginal regardless of $M$. 

\subsection{Quantum-Enhanced Distributed FOG Sensitivity with Constrained Total Fiber Length}
\label{sec:FixedFiberLength}

Gyroscopes are often employed in navigation or inertial tracking contexts with strict requirements on the total allowable payload and/or a maximum size restriction for the device. For a FOG with a fixed radius, these weight and size constraints will dictate a maximum total fiber length $L$ allowed for the gyroscope, assuming that fiber couplers, beamsplitters, and integrated optical sources and detectors account for a relatively small footprint. For a distributed FOG with $M$ identical interferometers, the fiber length allocated to each will be $L/M$. For each of the three distributed FOG modalities depicted in Fig. \ref{fig:distributedFOG}, the design parameter $M$ can be tuned to optimally balance the input coherent state amplitude $\alpha=\sqrt{Mn_v}$, the fiber transmissivity $\eta=10^{-b(L/M)/10}$, and the number of area-contributing fiber loops for each interferometer $m=L/(2\pi r M)$. 

Using Eq.~\ref{eq:DistributedVariance} for Design D and solving $\partial\expval*{\Delta\tilde{\Omega}_{\textrm{D}}^2}/\partial M=0$, the optimal number of distributed interferometers depends on the fiber length constraint $L$ and is given by
\begin{equation}
M_{\textrm{opt}}=\frac{bL\ln(10)}{10},
\label{eq:DistributedMopt}
\end{equation}
Consequently, the optimized estimator variance is
\begin{equation}
\expval*{\Delta\tilde{\Omega}_{\textrm{D}}^2}\vert_{M=M_{\rm opt}}=\frac{b e \ln(10)}{10L}.
\label{eq:DistributedFOGvariance}
\end{equation}
For Design E, the equation $\partial\expval*{\Delta\tilde{\Omega}_{\textrm{E}}^2}/\partial M=0$ can be solved using Eq. \ref{eq:variance5} to find 
\begin{equation}
M_{\textrm{opt}}=\frac{bL\ln(10)}{10(1+\Lambda(N_s))},
\label{eq:EntanglementEnhancedMopt}
\end{equation}
and
\begin{equation}
\expval*{\Delta\tilde{\Omega}_{\textrm{E}}^2}\vert_{M=M_{\rm opt}}=\frac{b e^{1+\Lambda(N_s)} \ln(10)}{10L}.
\label{eq:EntanglementEnhancedFOGvariance}
\end{equation}
A similar optimization for Design P does not lend to analytical solutions and can be performed numerically. The optimal multiplexing configuration for each design can be obtained by rounding $M_{\rm opt}$ either down or up to the nearest positive integer, whichever yields the lower estimator variance.

\begin{figure}[tbp]
	\centering
	\includegraphics[width=\linewidth]{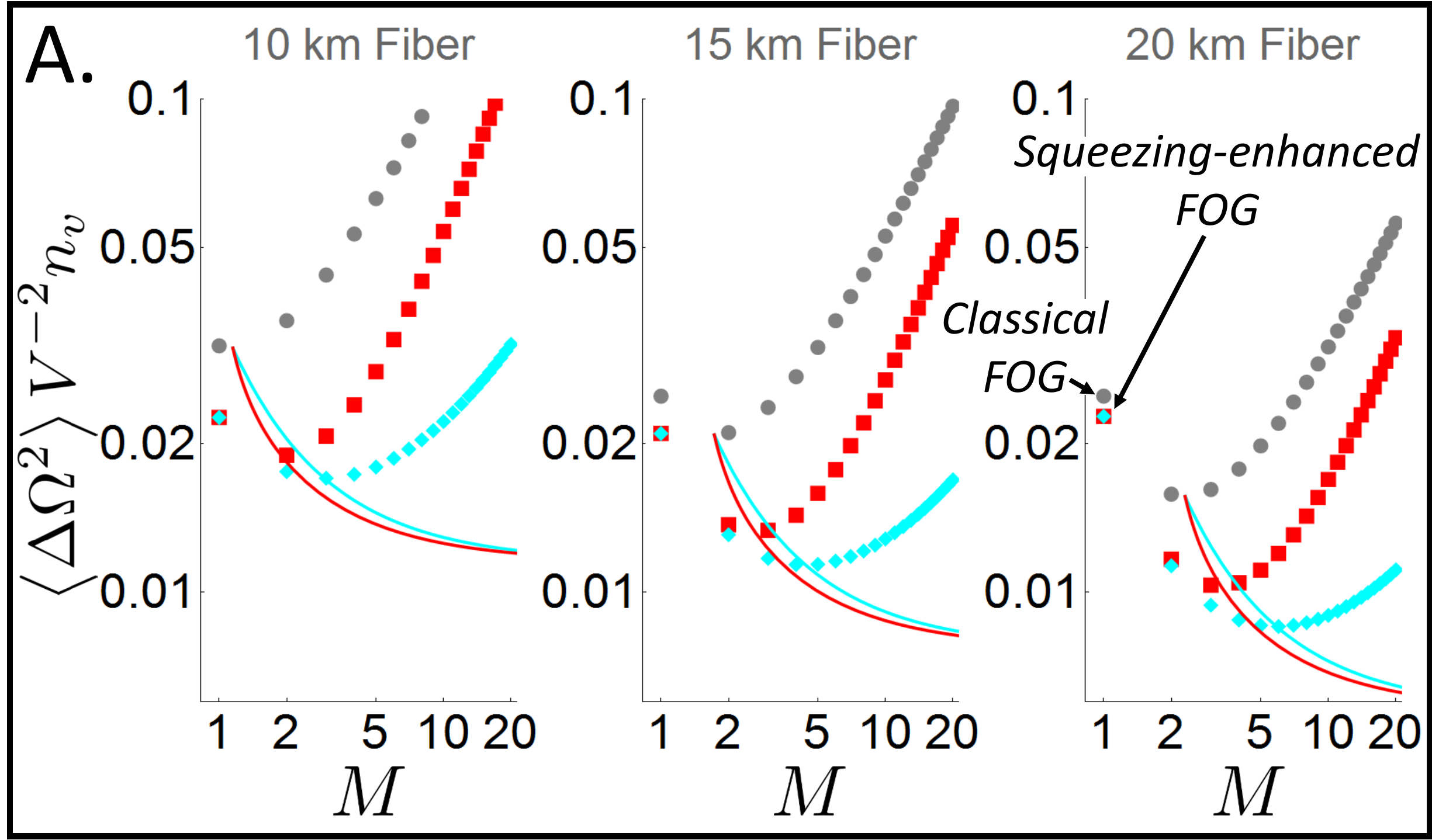}
	\caption{Normalized rotation estimate variance as a function of $M$ for Design D (gray), Design P FOG (red), and Design E (cyan) with given total fiber length constraints. In each case, $b=0.5$ dB loss/km of fiber. Design E results use 10 dB of squeezing. For the Design P results, the mean photon number $N_s$ from all injected SVs is equal to that of a single-mode SV with 10 dB squeezing. The descending solid parametric curves give rotation estimate variances with optimized $M$.}
	\label{fig:Fixed_Fiber_Length}
\end{figure}

The effect of this tradeoff on rotation sensitivity is visualized in Fig. \ref{fig:Fixed_Fiber_Length}A for three fixed total fiber lengths and 10 dB of squeezing. As the available fiber length is increased, the sensitivity achieved by all three distributed FOG designs improves, as expected. We find that adding one or two sensors will only improve Design D over Design C in the case of very long fiber lengths, whereas Designs P and E obtain a more robust benefit in sensitivity from multiplexing several interferometers. Design E reaches the best rotation sensitivity for a given mean photon number from squeezed light $N_s$, with an optimal number of interferometers $M_{\rm opt}$ that increases both with the total fiber length and with the squeezing but remains a reasonable $M_{\rm opt}\approx 5$ for 10 dB of squeezing. As in Fig. \ref{fig:BarChart}, the sensitivity of Design P is worse than that of the entanglement-enhanced FOG unless the per-mode squeezed photon number $n_s$ is constrained, in which case it matches that of Design E at the cost of $M$ single-mode squeezers and additional pump energy.

\section{Comparison of Quantum-Enhanced and Classical FOG Sensitivity}
\label{sec:ClassicalComparison}

While none of the FOG modalities described here exhibit a Heisenberg scaling advantage in the presence of fiber loss, quantum-enhancement can still enable notable constant factor improvements in sensor performance. In this section, we report the rotation sensitivity ratios $R_{\textrm{S}}=\expval*{\Delta\tilde{\Omega}_{\textrm{S}}^2}/\expval*{\Delta\tilde{\Omega}_{\textrm{C}}^2}$, $R_{\textrm{P}}=\expval*{\Delta\tilde{\Omega}_{\textrm{P}}^2}/\expval*{\Delta\tilde{\Omega}_{\textrm{D}}^2}$ and $R_{\textrm{E}}=\expval*{\Delta\tilde{\Omega}_{\textrm{E}}^2}/\expval*{\Delta\tilde{\Omega}_{\textrm{D}}^2}$ of Designs S, P and E against the corresponding classical FOGs in order to highlight the advantages of CV entanglement for FOG sensitivity. We stress that the baseline comparison for Designs P and E is Design D, so the $M$-fold reduction in estimator variance gained from increased laser power is canceled out in $R_{\textrm{P}}$ and $R_{\textrm{E}}$. As a result, these sensitivity ratios isolate the benefits arising from quantum-enhancement.

\subsection{Sensitivity Ratios with Fixed Fiber Length}

\begin{figure}[tbp]
	\centering
	\includegraphics[width=\linewidth]{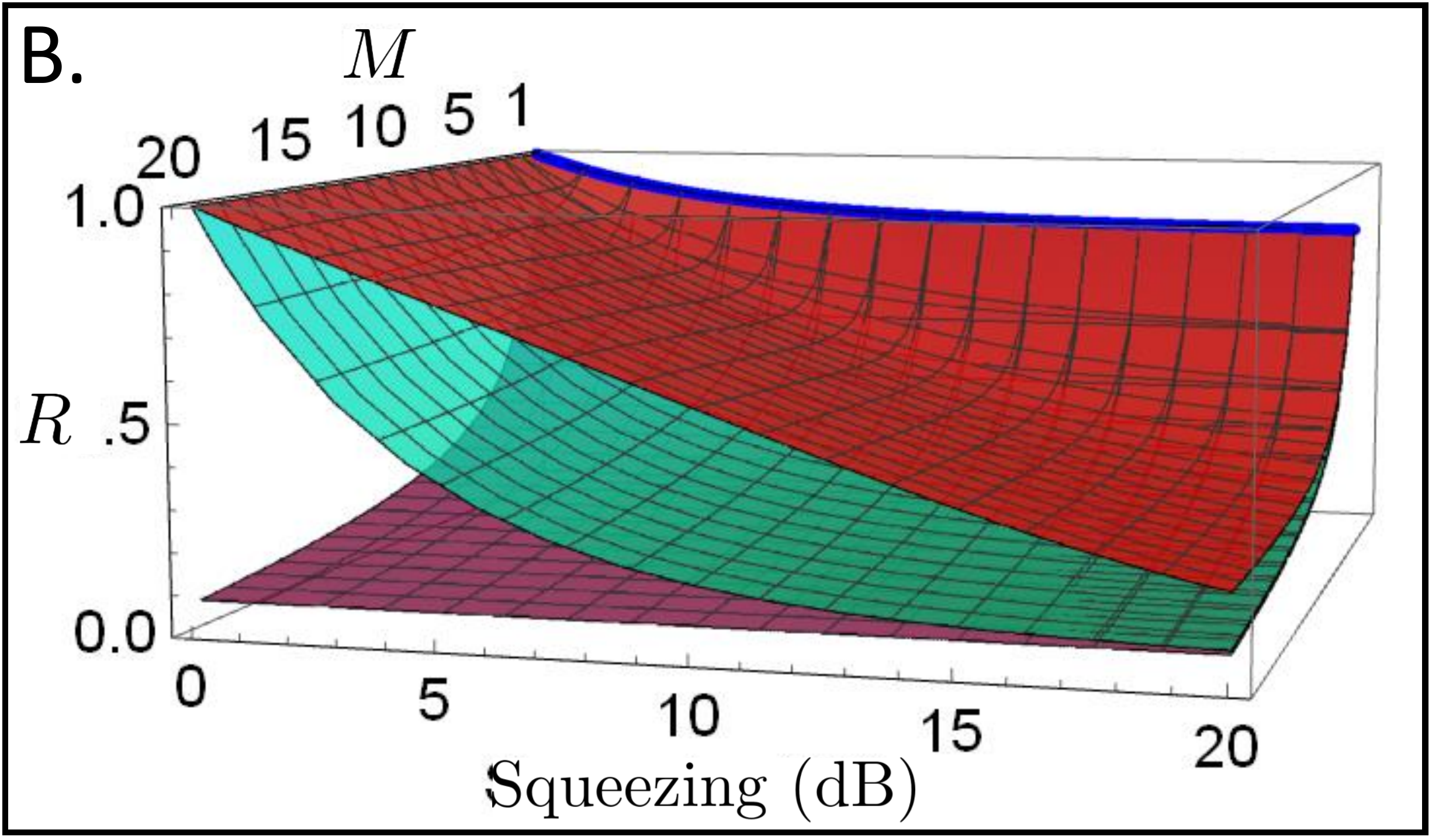}
	\caption{Sensitivity ratios $R_{\rm S}$ (blue parametric curve), $R_{\rm P}$ (red surface) and $R_{\rm E}$ (cyan surface) as a function of squeezing and number of interferometers for $b=0.5$ dB loss/km fiber and total fiber length constraint $L=15\textrm{ km}$. For the $R_{\rm P}$ results, the mean photon number $N_s$ from all injected squeezed states is equal to that of a single-mode SV state with squeezing given by the x-axis. The purple surface shows $1-\eta=1-10^{-b(L/M)/10}$ for reference.}
	\label{fig:3DPlot}
\end{figure}

As an example, Fig. \ref{fig:3DPlot} shows the three sensitivity ratios $R_{\rm S}$, $R_{\rm P}$, and $R_{\rm E}$ for the realistic scenario with a total fiber length requirement of $L=15$ km. While squeezing with a single interferometer (Design S) provides some improvement over a classical FOG (Design C), adding multiple distributed sensors augments the benefits of quantum-enhancement. In particular, the greatest performance gap between our Design E and the Design P occurs with 5 or more multiplexed interferometers and single-mode squeezing of 5--15 dB, which are reasonable design parameters for a next-generation quantum sensing system. Additionally, while the high-squeezing behavior $\lim_{N_s\to\infty}\{R_{\rm S}\}=\lim_{N_s\to\infty}\{R_{\rm P}\}=\lim_{N_s\to\infty}\{R_{\rm E}\}=1-\eta$ is recovered, Design E converges to this performance with a much lower energy requirement for squeezed light compared with Design P.

\subsection{Sensitivity Ratios for Optimized CV Entanglement-Enhanced FOGs}

Finally, we focus on our CV entanglement-enhanced FOG design and analytically calculate the constant-factor sensitivity ratio for optimized sensor configurations under two scenarios: when the total fiber length is not constrained and when it is constrained to a fixed length. In the unconstrained case, $R_{\rm E}\vert_{L=L_{\rm opt}}=\expval*{\Delta \tilde{\Omega}_{\textrm{E}}^2}\vert_{L=L_\textrm{opt}}/\expval*{\Delta \tilde{\Omega}_{\textrm{D}}^2}\vert_{L=L_\textrm{opt}}$ gives the ratio between the sensitivities of Design E (Eq. \ref{eq:Omega_ELopt}) and Design D (Eq. \ref{eq:Omega_DLopt}), where each estimator variance is calculated with optimized fiber length. Interestingly, this sensitivity ratio is found to be
\begin{equation}
R_{\textrm{E}}\vert_{L=L_\textrm{opt}}=\frac{2e^{\Lambda(N_s)}}{2+\Lambda(N_s)},
\label{eq:Rsqueezing}
\end{equation}
which depends on neither $N_v$, $b$, nor $M$ but is determined solely by the energy in the input SV state. In particular, since $R_{\textrm{E}}\vert_{L=L_\textrm{opt}}$ is the same for all values of $M$ including $M=1$, it is equivalent to the sensitivity ratio $R_{\textrm{S}}\vert_{L=L_\textrm{opt}}$ for a single-mode squeezing-enhanced FOG. In the high-squeezing limit, the sensitivity ratio converges to a lower bound of
\begin{equation}
\lim_{N_s\to\infty}\{R_{\textrm{E}}\vert_{L=L_\textrm{opt}}\}=\frac{2e^{W(-2/e^2)}}{2+W(-2/e^2)}=0.836,
\label{eq:limRsqueezing}
\end{equation} 
which sets the best possible constant-factor improvement in sensitivity with unconstrained total fiber length.

When the total fiber length $L$ is constrained to a fixed value, $R_{\textrm{E}}|_{M=M_{\textrm{opt}}}=\expval*{\Delta \tilde{\Omega}_{\textrm{E}}^2}\vert_{M=M_\textrm{opt}}/\expval*{\Delta \tilde{\Omega}_{\textrm{D}}^2}\vert_{M=M_\textrm{opt}}$ gives the ratio between the sensitivities of Designs E (Eq.~\ref{eq:EntanglementEnhancedFOGvariance}) and D (Eq.~\ref{eq:DistributedFOGvariance}). The sensitivity ratio 
\begin{equation}
R_{\textrm{E}}|_{M=M_{\textrm{opt}}}=e^{\Lambda(N_s)}
\label{eq:RE}
\end{equation}
depends only on $N_s$, not on $N_v$, $b$, nor $L$. Since these optimized sensitivities require $M_{\rm opt}\geq1$, this advantage is not achieved by Design S in general and requires the use of CV entanglement.  Remarkably, the high-squeezing limit gives the simple lower bound  
\begin{equation}
\lim_{N_s\to\infty}\{R_{\textrm{E}}|_{M=M_{\textrm{opt}}}\}=\frac{1}{e}\approx0.368,
\label{eq:RE_limit}
\end{equation}
the best constant-factor sensitivity improvement of our entanglement-enhanced FOG design over the corresponding classical system with a total fiber length constraint. 

\begin{figure}[tbp]
	\centering
	\includegraphics[width=\linewidth]{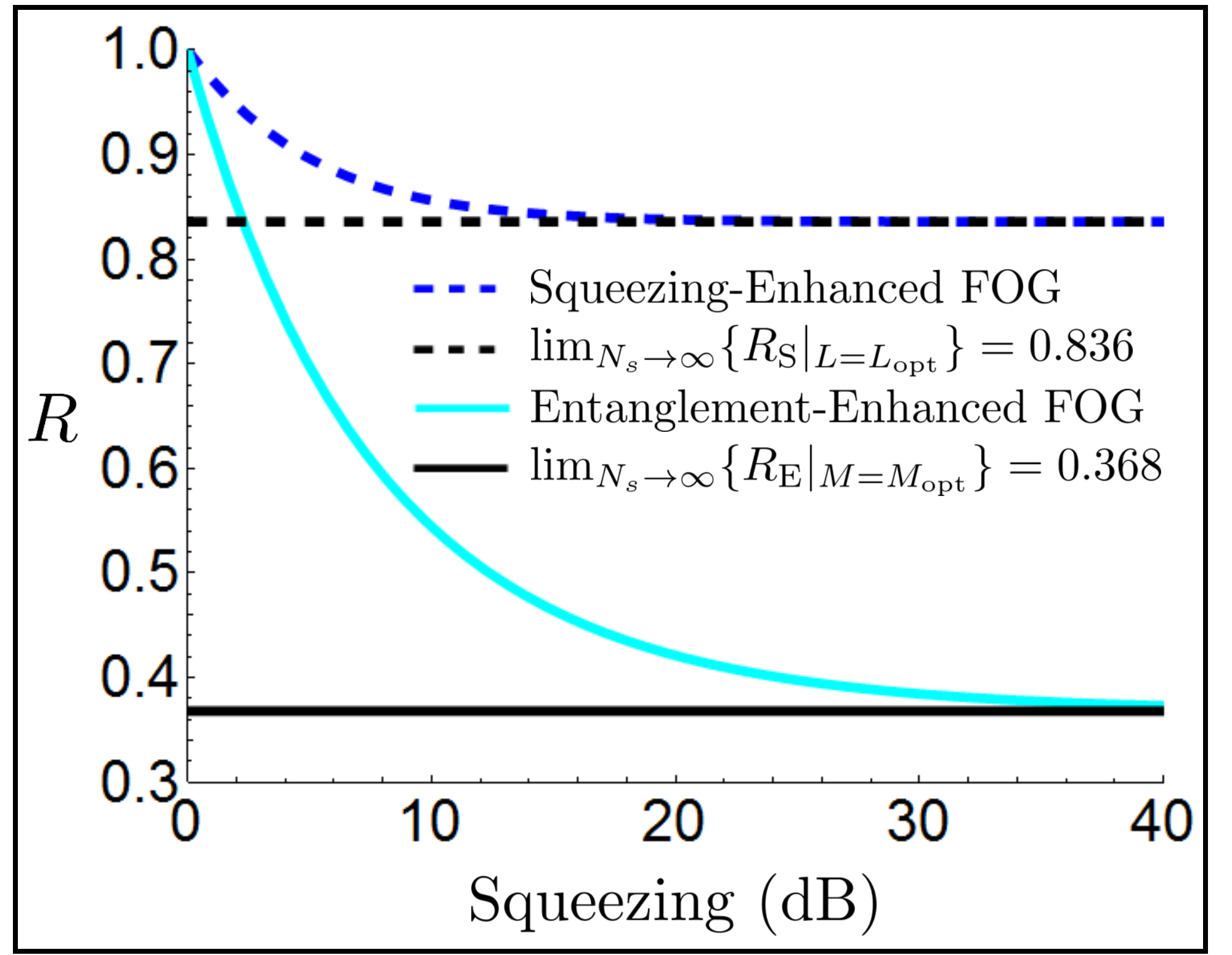}
	\caption{The ratio $R$ between angular-velocity estimator variances for quantum-enhanced and classical FOGs as a function of single-mode squeezing. Design S and Design E with unconstrained fiber length achieve a best sensitivity ratio of $0.836$. Design E achieves the minimum ratio of $1/e\approx0.368$ under a fixed total fiber length constraint.}
	\label{fig:ClassicalComparison}
\end{figure}

Fig. \ref{fig:ClassicalComparison} shows the sensitivity ratios under these two optimizations as a function of squeezing, and Table \ref{tab:squeezingConstraint} gives the constant-factor improvements (defined as the inverse of the sensitivity ratio) achieved by Design E. With unconstrained total fiber length, when Design E and Design S achieve equal sensitivity ratios, the lower bound on the sensitivity ratio is nearly saturated with a currently realizable 10--15 dB of squeezing. However, even the best constant-factor improvement of 1.196 is negligible, indicating that the system demonstrated in Ref.~\cite{Mehmet2010} has limited practical potential for improving rotation sensitivity. 

On the other hand, under the realistic constraint of fixed fiber length with an optimized number of interferometers, an appreciable improvement is possible, as the constant-factor improvement for our Design E over Design D in this case is upper bounded by $e=2.718$. CV entanglement is required to access this gain in performance, which holds regardless of the total fiber length requirement or the characteristic fiber loss. The maximum improvement is achieved at infinite squeezing, but most of this gain is attainable using 10--15 dB of squeezing (Table \ref{tab:squeezingConstraint}). As a result, our CV entanglement-enhanced FOG design could provide a meaningful quantitative benefit with currently feasible squeezing technology.
%Lossy homodyne?

\begin{table}[tbp]
	\centering
	\begin{tabular}{ |cccccc| } 
		\hline
		dB squeezing & 5 & 10 & 15 & 20 & $\infty$\\ 
		\hline
		\rule{0pt}{3ex}$R_{\rm E}\vert_{L=L_{\textrm{opt}}}^{-1}=R_{\rm S}\vert_{L=L_{\textrm{opt}}}^{-1}$ & 1.116 & 1.168 & 1.187 & 1.193 & 1.196\\ 
		\rule{0pt}{3ex}$R_{\rm E}\vert_{M=M_{\textrm{opt}}}^{-1}$ & 1.435 & 1.837 & 2.154 & 2.375 & 2.718\\
		\hline
	\end{tabular}
	\caption{Constant-factor improvements in rotation sensitivity for the entanglement-enhanced FOG compared with a distributed classical FOG under a total fiber length optimization and a total fiber length constraint with optimized number of interferometers.}
	\label{tab:squeezingConstraint}
\end{table}

\section{Discussion}
\label{sec:Discussion}
While none of the sensor designs evaluated here achieve Heisenberg-limited scaling in sensing performance, we have shown that FOG rotation sensitivity can be boosted by useful constant factors using CV squeezed light and quantum entanglement. Our analysis considers single-mode squeezing, optical gyroscope, and homodyne detection technologies that are possible with current technology. In fact, we have shown that additional improvements in single-mode squeezing capabilities would yield diminishing returns in rotation sensitivity. We conclude that an entanglement-enhanced FOG is a promising near-term technology that could both demonstrate the benefits of quantum sensing and potentially provide useful improvements to high-sensitivity inertial tracking systems under practical, real-world constraints such as size or payload limits.

A key advantage of our CV entanglement-enhanced FOG design is the need for only one single-mode squeezer. While a squeezing device could add to the total weight of a FOG setup, the ability to generate squeezed vacuum on chip would reduce the footprint compared with the optical fiber, although potentially at the cost of increased fiber coupling loss. An additional possibility could be generation of squeezed light inside of a cavity or optical fiber to entirely eliminate coupling. Finally, non-reciprocal phase accumulation due to scattering~\cite{Cutler1980}, polarization non-reciprocity, and nonlinear Kerr effects~\cite{Lefevre2014f} can be a significant source of error for FOG technologies and should be taken into account in future work.

The role of distributed sensing could be impactful for future technologies including FOGs. Indeed, entanglement could improve the performance of a FOG in inertial sensing contexts such as terrestrial, naval, aircraft and space-based navigation. Additionally, for coordinated movements with distant FOG systems, pre-shared entanglement could be used to register individual gyroscopes to each other. Experimental demonstration of our schemes is within reach of current technology, and we anticipate that our work will open an avenue to strategic and commercial applications.

\acknowledgments
This work was supported by the Office of Naval Research (ONR) under a 6.2 Quantum Information Science (QIS) project funded to University of Arizona, under contract number: N00014-19-1-2189. The authors acknowledge valuable discussions with Zheshen Zhang, Linran Fan, and Khanh Kieu.

\bibliographystyle{unsrt}
\bibliography{Fiber-Optic-Gyroscopes-Quantum_FOG}
\appendix
\section{Conjugate Phase Sensing Theory}
\label{apx:conjugage_phase_sensing}

Here, we derive the classical output statistics of a quantum-noise-limited, lossless homodyne measurement when the two arms of a conjugate-phase-sensing MZI are symmetrically affected by pure optical loss with transmissivity $\eta$ on each arm. In a conjugate phase sensing setup, such as a FOG of Designs C or S, the optical field operators $\hat{a}_{\textrm{in}}$ and $\hat{b}_{\textrm{in}}$ evolve through the system via unitary dynamics in the Heisenberg picture. The unitary transformation applied by the quantum circuit in the inset of Fig. \ref{fig:FOG} is
\begin{equation}
\begin{aligned}
U(\phi)&=\frac{1}{\sqrt{2}}
\begin{bmatrix}
1 & 1\\
-1 & 1
\end{bmatrix}
\begin{bmatrix}
e^{-i\phi} & 0 \\
0 & e^{i\phi}
\end{bmatrix}
\frac{1}{\sqrt{2}}
\begin{bmatrix}
1 & 1\\
1 & -1
\end{bmatrix} \\
&=
\begin{bmatrix}
\cos(\phi) & -i\sin(\phi) \\
i\sin(\phi) & -\cos(\phi)
\end{bmatrix}
\end{aligned}
\label{eq:unitary}
\end{equation}
where $i=\sqrt{-1}$. Eq.~\ref{eq:unitary} holds for both discrete-variable or continuous-variable quantum optical states.

Fiber loss will factor into any fiber-based sensor. Assuming symmetric loss processes that impose equal transmission coefficients $\eta$ on the two opposing paths around the fiber coil, we model loss with the pure-loss channel 
\begin{equation}
\Phi_{\eta}\bigg(
\begin{bmatrix}
\hat{a}_{\textrm{in}} \\
\hat{b}_{\textrm{in}}
\end{bmatrix}\bigg)
=
\begin{bmatrix}
\sqrt{\eta}\hat{a}_{\textrm{in}} + \sqrt{1-\eta}\hat{e}_a \\
\sqrt{\eta}\hat{b}_{\textrm{in}} + \sqrt{1-\eta}\hat{e}_b
\end{bmatrix},
\label{eq:pure-loss}
\end{equation}
where $\hat{e}_k$ signifies an ancillary vacuum mode that is associated with the optical mode $\hat{k}_{\textrm{in}}$. For any Gaussian state, a pure loss channel with equal loss for each mode commutes with any passive unitary operation, so we can act the loss channel $\Phi_{\eta}$ directly on the input modes $\hat{a}_{\textrm{in}}$ and $\hat{b}_{\textrm{in}}$. The output field operators for a FOG become
\begin{equation}
\begin{aligned}
\begin{bmatrix}
\hat{a}_{\textrm{out}}\\
\hat{b}_{\textrm{out}}
\end{bmatrix}
=&U(\phi)\Phi_{\eta}\bigg(
\begin{bmatrix}
\hat{a}_{\textrm{in}}\\
\hat{b}_{\textrm{in}}
\end{bmatrix}\bigg)
\\ =&
\begin{bmatrix}
\begin{aligned}
\cos&(\phi)(\sqrt{\eta}\hat{a}_{\textrm{in}}+\sqrt{1-\eta}\hat{e}_a) \\
-&i\sin(\phi)(\sqrt{\eta}\hat{b}_{\textrm{in}}+\sqrt{1-\eta}\hat{e}_b)
\end{aligned}\\
\begin{aligned}
i\sin&(\phi)(\sqrt{\eta}\hat{a}_{\textrm{in}}+\sqrt{1-\eta}\hat{e}_a)\\
-&\cos(\phi)(\sqrt{\eta}\hat{b}_{\textrm{in}}+\sqrt{1-\eta}\hat{e}_b)
\end{aligned}
\end{bmatrix}.
\end{aligned}
\label{eq:output_operators}
\end{equation}

A homodyne measurement on one quadradure of one of the two output modes is sufficient to estimate $\phi$ for given input states $\ket{\psi_a}$ and $\ket{\psi_b}$. If the homodyne measurement is quantum-noise-limited, the classical random variable $\tilde{b}_{\textrm{out}}$ will inherit the statistics of $\textrm{Im}[\hat{b}_{\textrm{out}}]$. Its mean is
\begin{equation}
\begin{aligned}
\expval*{\tilde{b}_{\textrm{out}}}=&\expval*{\textrm{Im}[\hat{b}_{\textrm{out}}]} \\
=& \sin(\phi)\big(\sqrt{\eta}\expval*{\textrm{Re}[\hat{a}_{\textrm{in}}]} +\sqrt{1-\eta}\expval*{\textrm{Re}[\hat{e}_a]}\big) \\
&-\cos(\phi)\big(\sqrt{\eta}\expval*{\textrm{Im}[\hat{b}_{\textrm{in}}]}+\sqrt{1-\eta}\expval*{\textrm{Im}[\hat{e}_b]}\big) \\
=& \sqrt{\eta}\big(\sin(\phi)\expval*{\textrm{Re}[\hat{a}_{\textrm{in}}]}-\cos(\phi)\expval*{\textrm{Im}[\hat{b}_{\textrm{in}}]}\big),
\end{aligned}
\label{eq:homodyne_mean_apx}
\end{equation}
and its variance is
\begin{equation}
\begin{aligned}
\expval*{\Delta\tilde{b}_{\textrm{out}}^2}=& 
\expval*{
	\textrm{Im}[\hat{b}_{\textrm{out}}]^2
}- \expval*{\textrm{Im}[\hat{b}_{\textrm{out}}]}^2 \\
=& \sin(\phi)^2\big(\eta\expval*{\textrm{Re}[\hat{a}_{\textrm{in}}]^2} + (1-\eta)\expval*{\textrm{Re}[\hat{e}_a]^2}\big) \\
&+\cos(\phi)^2\big(\eta\expval*{\textrm{Im}[\hat{b}_{\textrm{in}}]^2}+(1-\eta)\expval*{\textrm{Im}[\hat{e}_b]^2}\big)\\
& -2\eta\sin(\phi)\cos(\phi)\expval*{\textrm{Re}[\hat{a}_{\textrm{in}}]}
\expval*{\textrm{Im}[\hat{b}_{\textrm{in}}]} \\
&-\eta\big(\sin(\phi)^2\expval*{\textrm{Re}[\hat{a}_{\textrm{in}}]}^2
+\cos(\phi)^2\expval*{\textrm{Im}[\hat{b}_{\textrm{in}}]}^2\big)\\
&+2\eta\sin(\phi)\cos(\phi)\expval*{\textrm{Re}[\hat{a}_{\textrm{in}}]}\expval*{\textrm{Im}[\hat{b}_{\textrm{in}}]}\\
=& \eta\big(\sin(\phi)^2\expval*{\Delta\textrm{Re}[\hat{a}_{\textrm{in}}]^2}\\
&+\cos(\phi)^2\expval*{\Delta\textrm{Im}[\hat{b}_{\textrm{in}}]^2}\big)+\frac{1-\eta}{4}.
\end{aligned}
\label{eq:homodyne_variance_apx}
\end{equation}
Quantum vacuum state quadratures exhibit zero mean field ($\expval*{\textrm{Re}[\hat{e}_a]}=\expval*{\textrm{Im}[\hat{e}_b]}=0$) but nonzero variance ($\expval*{\Delta\textrm{Re}[\hat{e}_a]^2}=\expval*{\Delta\textrm{Im}[\hat{e}_b]^2}=1/4$), the latter being a direct consequence of the quantum uncertainty principle. %hbar??????

\section{Distributed Conjugate Phase Sensing}
\label{apx:distributed_conjugate_phase_sensing}

A distributed FOG with separable inputs to each interferometer (e.g., Designs D and P) can be thought of as a series of conjugate phase sensors, where the identical output modes can be combined on a balanced beamsplitter array to coalesce all of the information into one output mode $\hat{b}'_{\textrm{out},1}$. After recombination, this mode is related to the input modes: 
\begin{equation}
\begin{aligned}
\hat{b}'_{\textrm{out},1}=&\expval*{\textrm{Im}[\hat{b}'_{\textrm{out},1}]} \\
=&i\sin(\phi)(\sqrt{\eta}\hat{a}'_{\textrm{in},1}+\sqrt{1-\eta}\hat{e}'_{a,1}) \\
&-\frac{\cos(\phi)}{\sqrt{M}}\sum_j^M(\sqrt{\eta}\hat{b}_{\textrm{in},j}+\sqrt{1-\eta}\hat{e}_{b,j}).
\end{aligned}
\label{eq:product_operators}
\end{equation}  
The homodyne measurement on the imaginary quadrature of mode $\hat{b}'_{\textrm{out},1}$ then has mean
\begin{equation}
\begin{aligned}
\expval*{\tilde{b}'_{\textrm{out},1}}=&\sin(\phi)(\sqrt{\eta}\expval*{\textrm{Re}[\hat{a}'_{\textrm{in}_1}]}+\sqrt{1-\eta}\expval*{\textrm{Re}[\hat{e}'_{a,1}]}) \\
&-\frac{\cos(\phi)}{\sqrt{M}}\sum_j^M(\sqrt{\eta}\expval*{\textrm{Im}[\hat{b}_{\textrm{in},j}]}+\sqrt{1-\eta}\expval*{\textrm{Im}[\hat{e}_{b,j}]})\\
=& \sqrt{\eta}\big(\sin(\phi)\expval*{\textrm{Re}[\hat{a}'_{\textrm{in}_1}]} -\frac{\cos(\phi)}{\sqrt{M}}\sum_j^M\expval*{\textrm{Im}[\hat{b}_{\textrm{in},j}]}\big),
\end{aligned}
\label{eq:distributed_homodyne_mean}
\end{equation}
\vspace{50pt}
and variance
\begin{equation}
\begin{aligned}
\expval*{\Delta\tilde{b}_{\textrm{out},1}^{\prime2}}=&\expval*{
	\textrm{Im}[\hat{b}'_{\textrm{out},1}]^2
}- \expval*{\textrm{Im}[\hat{b}'_{\textrm{out},1}]}^2 \\
=& \sin(\phi)^2\Big(\eta\expval*{\textrm{Re}[\hat{a}'_{\textrm{in},1}]^2} + \frac{1-\eta}{4}\Big)\\
&+\frac{\cos(\phi)^2}{M}\sum_j^M\Big(\eta\expval*{\textrm{Im}[\hat{b}_{\textrm{in},j}]^2}+\frac{1-\eta}{4}\Big)\\
&+ \frac{2\eta\cos(\phi)^2}{M}\sum_{j\neq k}^M \expval*{\textrm{Im}[\hat{b}_{\textrm{in},j}]}\expval*{\textrm{Im}[\hat{b}_{\textrm{in},k}]}\\
& -\frac{2\eta\sin(\phi)\cos(\phi)}{\sqrt{M}}\sum_j^M \expval*{\textrm{Re}[\hat{a}'_{\textrm{in},1}]}
\expval*{\textrm{Im}[\hat{b}_{\textrm{in},j}]} \\
&-\eta\sin(\phi)^2\expval*{\textrm{Re}[\hat{a}'_{\textrm{in},1}]}^2\\
&-\frac{\eta\cos(\phi)^2}{M}\sum_j^M\expval*{\textrm{Im}[\hat{b}_{\textrm{in},j}]}^2\\
&-\frac{2\eta\cos(\phi)^2}{M}\sum_{j\neq k}^M \expval*{\textrm{Im}[\hat{b}_{\textrm{in},j}]}\expval*{\textrm{Im}[\hat{b}_{\textrm{in},k}]}\\
& +\frac{2\eta\sin(\phi)\cos(\phi)}{\sqrt{M}}\sum_j^M \expval*{\textrm{Re}[\hat{a}'_{\textrm{in},1}]}
\expval*{\textrm{Im}[\hat{b}_{\textrm{in},j}]}\\
=& \eta\big(\sin(\phi)^2\expval*{\Delta\textrm{Re}[\hat{a}'_{\textrm{in},1}]^2}\\
&+\frac{\cos(\phi)^2}{M}\sum_j^M\expval*{\Delta\textrm{Im}[\hat{b}_{\textrm{in},j}]^2}\big)+\frac{1-\eta}{4}.
\end{aligned}
\label{eq:distributed_homodyne_variance}
\end{equation}

On the other hand, the injected SV in Design E cannot be written as a series of separable quantum states. The input/output field operator relationship in this case is given by
\begin{equation}
\begin{aligned}
\hat{b}'_{\textrm{out},1}=&i\sin(\phi)(\sqrt{\eta}\hat{a}'_{\textrm{in},1}+\sqrt{1-\eta}\hat{e}'_{a,1}) \\
&-\cos(\phi)(\sqrt{\eta}\hat{b}'_{\textrm{in},1}+\sqrt{1-\eta}\hat{e}'_{b,1}),
\end{aligned}
\label{eq:entangled_operators}
\end{equation}  
which has the exact same form as the single-interferometer case (Eq. \ref{eq:output_operators}). Since mode $\hat{b}'_{\textrm{in},1}$ is the only SV state and has a mean photon number of $N_s$, the situation is identical to the development for Design S (Section \ref{sec:quantumFOG}), and the same results for homodyne measurement statistics (Eqs. \ref{eq:homodyne_mean_apx} and \ref{eq:homodyne_variance_apx}) can be used, where $\tilde{b}_{\textrm{out}}$ should be replaced with $\tilde{b}'_{\textrm{out},1}$, $\hat{a}_{\textrm{in}}$ with $\hat{a}'_{\textrm{in},1}$, and $\hat{b}_{\textrm{in}}$ with $\hat{b}'_{\textrm{in},1}$.

\end{document}